\documentclass[prl,twocolumn,amsmath,amssymb]{revtex4}

\usepackage{graphicx}
\usepackage{dcolumn}
\usepackage{bm}

\begin{document}

\title{Vortex manipulation in a superconducting matrix with view on applications}

\author{M. V. Milo\v{s}evi\'{c}}
\email{milorad.milosevic@ua.ac.be}
\author{F. M. Peeters}

\affiliation{Departement Fysica, Universiteit Antwerpen,
Groenenborgerlaan 171, B-2020 Antwerpen, Belgium}

\date{\today}

\begin{abstract}
We show how a single flux quantum can be effectively manipulated in
a superconducting film with a matrix of {\it blind holes}. Such a
sample can serve as a basic memory element, where the position of
the vortex in a $k\times l$ matrix of pinning sites defines the
desired combination of $n$ bits of information ($2^n=k\cdot l$).
Vortex placement is achieved by strategically applied current and
the resulting position is read-out via generated voltage between
metallic contacts on the sample. Such a device can also act as a
controllable source of a nanoengineered local magnetic field for
e.g. spintronics applications.
\end{abstract}

\maketitle

Superconducting electronics has always been envisaged as a candidate
for futuristic applications, thanks to its low resistance, low
dissipation, and high current densities which allow for high
power/size ratios. In the last decade, enormous research efforts
have been delivered in the field of mesoscopic superconductivity,
where samples are comparable to the characteristic superconducting
length scales (coherence length $\xi$ and penetration depth
$\lambda$) and therefore exhibit pronounced quantum effects. The
revealed key dynamic effects include: the step-like resistance of
the superconducting elements as a function of applied current (zero
resistance - `resistive' state - normal state), and the
corresponding definition of two critical currents \cite{resstate};
very rich `phase-slip' phenomena \cite{psreview}; S- and N- shaped
I-V characteristics of superconducting stripes and wires
\cite{SNshape}; control of dynamic properties of the sample by
perforations or magnetic structuring \cite{enhpar}; `ratchet'
physics, using the mobility of vortices in applied current across
asymmetric pinning potentials \cite{ratchet}. The latter is the
current pivotal axis of the field of {\it fluxonics}, the research
area exploiting duality between electrons and superconducting flux
quanta in electromagnetic fields.
\begin{figure}[b]
\includegraphics[width=0.8\linewidth]{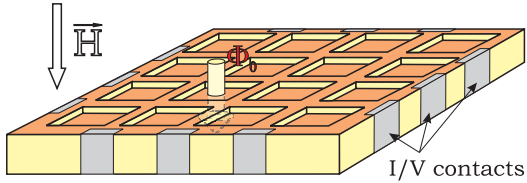}
\caption{\label{fig1} The oblique view of the sample: a
superconducting square (size $w\times w$ and thickness $d$) with a
$4\times4$ lattice of blind holes (each with size $a\times a$ and
thickness $d_b$). Shown direction of applied homogeneous magnetic
field $H$ is denoted positive.}
\end{figure}

In this Letter we show the use of fluxonics in a superconducting
matrix, i.e. the manipulation of a single vortex in a square sample
with arrays of blind holes (see Fig. \ref{fig1}). As we will show,
such a sample can act as a superconducting memory device, with
individually addressable memory cells and without restrictions on
read/write cycles. Additionally, the sample can act as a spatially
controllable field source, which is of use in nanoscale spintronics
and hybrid structures.

The concept of here presented devices is based on the following
electron-vortex analogies: (i) an electric current drives vortices
in the same manner as an electromagnetic field drives electrons
(Lorentz force), and (ii) moving vortices produce voltages similar
to mobile electrons producing electric currents. To characterize
this behavior, we use the suitably modified time-dependent
Ginzburg-Landau equation \cite{Kramer}
\begin{eqnarray}
\frac{u}{\sqrt{1+\Gamma^2|\psi|^2}} \left(\frac {\partial }{\partial
t} +{\rm i}\varphi +
\frac{\Gamma^2}{2}\frac{\partial|\psi|^2}{\partial t}
\right)\psi= \nonumber \\
=(\nabla - {\rm i} {\bf A})^2 \psi +(1-|\psi|^2)\psi + \frac{\nabla
d(x,y)}{d(x,y)}(\nabla - {\rm i} {\bf A}) \psi, \label{tdgl1}
\end{eqnarray}
coupled with the equation for the electrostatic potential
\begin{eqnarray}
\Delta \varphi = {\rm div}\left(\Im(\psi^*(\nabla-{\rm i}{\bf
A})\psi)\right). \label{tdgl2}
\end{eqnarray}
Note that the last term in Eq. (\ref{tdgl1}) accounts for the
variable thickness of the sample $d(x,y)$, while equations remain
averaged in the $z$-direction (i.e. uniform distribution of all
quantities is assumed across the sample thickness) \cite{bhgolib}.
In Eqs. (\ref{tdgl1}-\ref{tdgl2}), the distance is measured in units
of $\xi$, $\psi$ is scaled by its value in the absence of magnetic
field $\psi_{0}$, time by $\tau_{GL}=2T\hbar \big/\pi\psi_0^2$,
vector potential ${\bf A}$ by $c\hbar\big/ 2e\xi$, and the
electrostatic potential by $\varphi_0=\hbar\big/2e\tau_{GL}$.
$\Gamma=2\tau_E \psi_{0}/\hbar$ is directly proportional to
inelastic electron-collision time $\tau_E$, and equals 10 in the
present calculation, and parameter $u=5.79$ is taken from Ref.
\cite{Kramer}. The normal-metal leads where current is injected in
the sample (see Fig. \ref{fig1}) satisfy $-\nabla \varphi=j_i$,
where $j_i$ is the injected current density in units of
$j_0=c\Phi_0\big/8\pi^2\Lambda^2\xi$, with $\Lambda=\lambda^2/d$.
Edges of the samples are modeled by the Neumann boundary condition
at superconductor/vacuum interfaces. Here considered samples are
thin, and we may therefore neglect the screening of the applied
magnetic field ${\bf H}=(0,0,H)$, and take ${\bf
A}$=($\frac{1}{2}Hy$,$-\frac{1}{2}Hx$,0) in Eq. (\ref{tdgl1}).
Nevertheless, this still allows us to eventually calculate the
magnetic response of the sample, using ${\bf
j}_s=\Im(\psi^*\nabla\psi)-|\psi|^2{\bf A}$ to obtain the
supercurrent density in the sample.

The key idea of the superconducting memory is to have the position
of a single vortex represent one combination of several bits of
data. For example, to possibly store all combinations of four bits
of data, one needs $2^4=16$ logic states. To represent this in a
superconducting memory based on just one vortex, we need a sample
with 16 possible single-vortex states. We illustrate one possible
candidate in Fig. \ref{fig1}, where the vortex can be located in any
of the 4x4 blind holes\cite{blhole}. Because vortices in mesoscopic
superconductors generally favor central positions in the sample, due
to interaction with strong Meissner currents at sample edges, it is
necessary to nano-engineer such samples with sufficiently spaced,
large and deep holes such that each one of them is capable to pin a
vortex. In what follows, we focus on a sample with size $w\times
w=32\xi\times32\xi$ and thickness $d=1\xi$, with uniformly
distributed square blind holes of size $a=4\xi$ and depth $0.9\xi$
(i.e. with bottom thickness $d_b=0.1\xi$, see Fig. \ref{fig1}).

\begin{figure}[t]
\includegraphics[width=\linewidth]{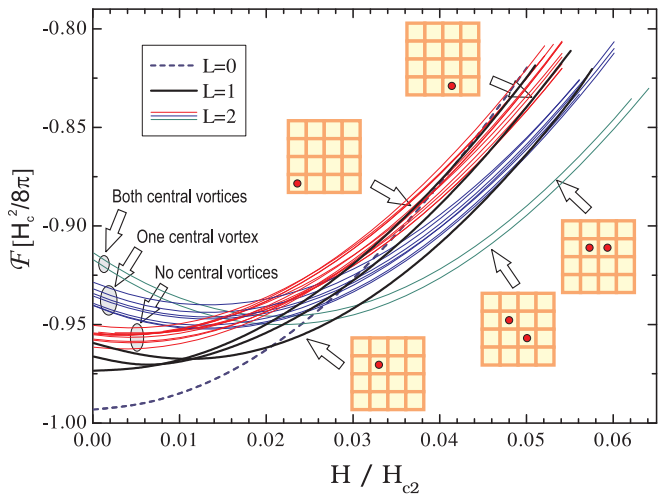}
\caption{\label{fig2} The free energy of stable vortex states in the
4x4 superconducting matrix (with up to two vortices). The complexity
of the energy landscape grows with vorticity. Insets depict the
position of vortices in the sample.}
\end{figure}
In Fig. \ref{fig2}, we show the behavior of the above described
superconducting sample in applied perpendicular magnetic field. At a
given value of the magnetic field, we search through the Gibbs
energy landscape for all stable states, with or without vortices.
The free energy is calculated using
$\mathcal{F}=\frac{H_c^2}{8\pi}\frac{d(x,y)}{V}\int_V |\psi|^4 dV$,
where $V$ is the sample volume). In Fig. \ref{fig2}, we show the
energy levels obtained for states with vorticity $L\leq 2$. Although
there are 16 stable $L=1$ states (further denoted by $(m,n)$, with a
vortex located in $m$-th hole in $x$-, and $n$-th hole in
$y$-direction), only three energy levels exist (two quadruple, and
one octuple degenerate) due to the four-fold sample symmetry (see
insets in Fig. \ref{fig2}). In the case of two vortices, we find 21
distinct energy levels for 120 possible states \cite{foot1}. In our
memory cell, we aim to use a single vortex scenario, although $L=2$
state offers storage of additional 120 bit-combinations. The reason
is that two vortices are much more difficult to control
simultaneously in the present concept, but this possibility should
not be entirely disregarded.
\begin{figure}[b]
\includegraphics[width=\linewidth]{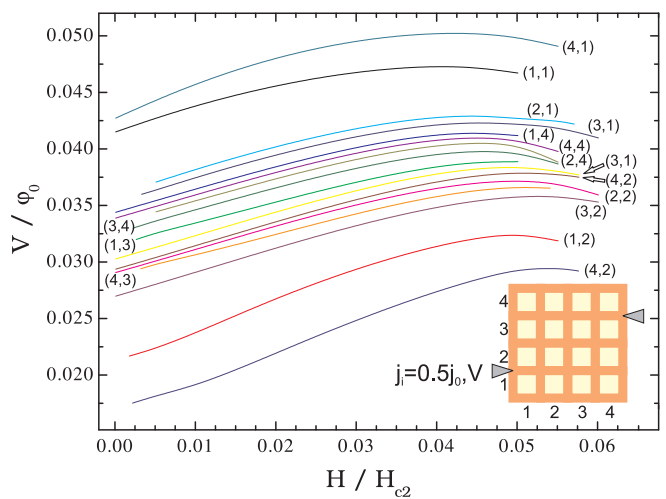}
\caption{\label{fig3} Measured voltage diagonally across the sample,
for small test current applied, for all possible positions of a
single vortex in the matrix (i.e. position read-out)\cite{foot2}.}
\end{figure}

If each of the possible positions of one vortex in our memory cell
is to represent a combination of bits, we must first enable
successful readout of those states, and the degeneracy of energy
levels in Fig. \ref{fig2} is not helpful. For that reason, we
suggest injection of a weak test current in the sample, and that in
a diagonal direction across the cell. In that case, the vortex,
regardless of its position, will experience a Loretzian force
perpendicular to the applied current, but its response will depend
on its exact position. For example, the energy-degenerate vortex
states $(1,1)$ and $(4,4)$ will feel the drive towards the interior
of the sample and out of the sample, respectively, and their energy
levels must split. To detect these subtle differences, we propose
the measurement of voltage between the current leads. Since we use
the normal-metal leads, the normal current survives in the
superconducting sample up to certain characteristic length. However,
the length over which the non-equilibrium quasiparticles can exist
in the sample $\mathcal{L}=\sqrt{D\tau_E}$ ($D$ being the diffusion
constant \cite{Denis1}) is typically larger than the size of a
mesoscopic sample. For that reason, in our memory cell the normal
current can reach the lead across the sample, and a finite voltage
can be detected \cite{foot3}. The measured voltage due to injected
quasiparticles will depend on their entire path and Andreev
recombinations at the vortex core, and will therefore differ for
every position of the vortex in the matrix. We show this in Fig.
\ref{fig3}, as an evolution of the calculated voltage between the
leads as a function of applied field, for small injected current
$j_i$, and all 16 possible $L=1$ states in their full stability
range. Since indeed all 16 voltages clearly differ, this enables the
successful read-out of the vortex position.

\begin{figure}[t]
\includegraphics[width=\linewidth]{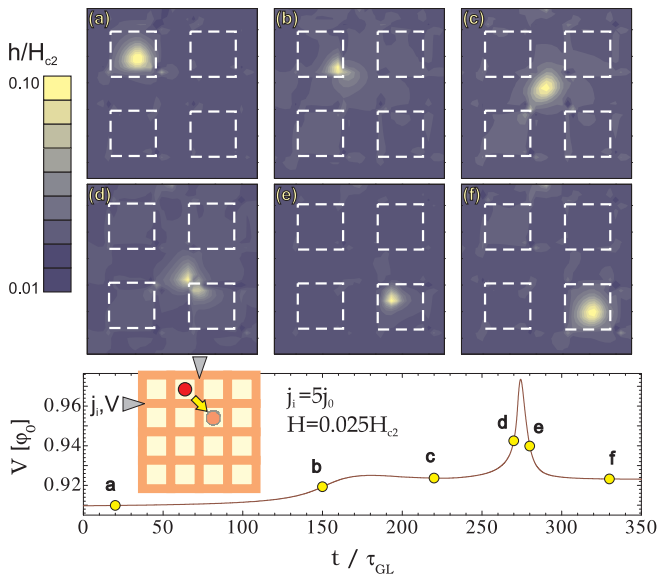}
\caption{\label{fig4} Measured voltage at the leads for current
injection\cite{foot4}, during the vortex manipulation in the cell
(desired positioning) as shown in the inset. (a-e) Zoomed-in
distribution of the magnetic field above the sample during the
operation.}
\end{figure}

In order to write the data in the memory cell, we must realize
single-vortex manipulation at nanoscale. For that we use the same
principle - the Lorentzian behavior of vortices in an applied
current. Sufficiently large current will be able to depin the vortex
from the residing hole and push it towards another location. Of
course, to move the vortex from one particular position to another,
one must apply the current {\it strategically}. For example, to move
the vortex from location $(2,4)$ to $(3,3)$, the current could be
applied between columns 2 and 3 to move the vortex `right', and
additionally between rows 3 and 4 to move the vortex down.
Alternatively, the latter two currents can be combined in one, as
shown in Fig. \ref{fig4}. Successful vortex hopping can be monitored
by measured voltage at the current leads, as it leaves a distinct
feature (maximum) in the voltage vs. time characteristics.

Fig. \ref{fig4} also shows the magnetic field profile under the
sample, resulting from a vortex in motion. Such localized and
moveable sources of magnetic field recently became of significant
technological relevance, thanks to the prediction of Berciu {\it et
al.} \cite{janko}. Namely, electronic and spin states in
dillute-magnetic-semiconductors (DMS) seem to be very responsive to
a non-homogeneous magnetic field, and can be trapped under a vortex
core in a DMS-superconductor bilayer. This also means that those
spins and charges can be manipulated by manipulating vortices, and
our device provides the needed control. Besides its potential for
spintronics, manipulation of the vortex position can also provide
controllable switching for magnetic cellular automata \cite{imre}.

Finally, we reflect on few potential problems of the device. Since
the vortex in our system is driven by an applied current, the
magnitude of the current is of outmost importance. Obviously, the
lowest current needed for the successful readout is determined by
the sensitivity of the voltage measurement, but can easily be in the
nA range. However, the current needed for the vortex manipulation
from one pin to the other strongly depends on the strength of the
pinning site. This justifies our choice of blind holes, since they
are able to hold the vortex, but not as strongly as a full
perforation. Additionally, blind holes enable direct visualization
of the vortex core, and testing of the device by low-temperature
magnetic \cite{magn} and tunneling \cite{STM} scanning probe
measurements. Depending on the size and depth of the hole, the
threshold current for vortex depinning has to be determined
\cite{foot5}. Used current in the device has to be close to the
minimal needed one, since a larger current may force the vortex to
overshoot the desired position, even leave the sample, and very
large injected current can cause phase-slippage, finite resistance
and dissipation \cite{resstate}. On a positive note, these error
scenarios are detectable, since each trapping of the vortex in a
hole causes minima, and vortex exit maxima in the measured voltage
versus time \cite{milokanda}. In summary, although latter issues and
its operation at low temperatures hamper the applicability of the
device, we have here demonstrated the proof of concept for a
single-vortex superconducting matrix with a high level of control.
This concept is verifiable in experiment, it has an intuitive
application as a superconducting memory, and can lead to further
developments of e.g. controllable nanoscale field sources for
applications in hybrid devices.

This work was supported by the Flemish Science Foundation (FWO-Vl),
the Belgian Science Policy (IAP), the ESF-NES and ESF-AQDJJ
networks.


\begin{thebibliography}{}
\bibitem{foot1} The size of the holes is insufficient to trap both vortices, otherwise the number of possible
states would be 136, with 24 distinct energy levels.
\bibitem{foot2} $\varphi_0$ is typically 50-100 $\mu$V in low-T$_c$ samples, and can be
significantly higher in high-T$_c$ materials. $j_0$ is approximately
in the $10^6-10^7$ A/cm$^2$ range, making the injected current in
this case c.a. 0.1-1 $\mu$A.
\bibitem{foot3} Therefore, even in the fully superconducting state, the sample exhibits
a non-zero resistance due to the metallic leads.
\bibitem{foot4} The Ginzburg-Landau time $\tau_{GL}$ is typically in the 1-10 ps
range at intermediate temperatures.
\bibitem{foot5} For an Al cell with given geometrical
parameters, we obtained a depinning current of $\approx10$ $\mu$A.
\end{thebibliography}
\end{document}